\documentclass[twocolumn,pra,showpacs,preprintnumbers,amsmath,amssymb]{revtex4}
\usepackage{color}
\usepackage{bm, graphicx, amsmath}
\usepackage[mathscr]{eucal}
\usepackage{subcaption}
\usepackage{float}

\begin{document}

\title{Interference and Measurement: Changing amplitude phase information to amplitude magnitude information}
\author{Pranjal Agarwal, Nada Ali, and Mark Hillery}
\affiliation{Department of Physics and Astronomy, Hunter College of the City University of New York, 695 Park Avenue, New York, NY 10065 USA \\ 
Physics Program, Graduate Center of the City University of New York, 365 Fifth Avenue, New York, NY 10016} 

\begin{abstract}
There are quantum procedures that encode the solutions to a problem in the phases of quantum amplitudes.  This happens in some quantum optimization algorithms in which the value of a function to be maximized or minimized is represented by a phase.  An example of this is the QAOA algorithm for the MaxCut problem in which one encodes the number of edges connecting the sets resulting from a partition of the vertices of a graph into phases of amplitudes of a quantum state.  Another is the minimum vertex cover problem in which the number of edges included in the cover is encoded in phases.  Here we want to see what can be done if we only use simple aspects of quantum mechanics, interference and measurement, to manipulate the magnitudes of the amplitudes whose phases encode the relevant information.  The idea is to use constructive interference to enhance the amplitudes that contain useful information and destructive interference to suppress those that do not.   We examine examples, both analytically and numerically.  We also show how the results of sequences of measurements can be used to gain information about the landscape of solutions.
\end{abstract}
\maketitle

\section{Introduction}
Suppose one is given a system in a quantum state in which useful information, for example, the solutions to a problem, are encoded in the phases of the amplitudes of the state when it is expressed in the computational basis.  This occurs is some quantum approaches to optimization problems in which there is a function that one would like to minimize of maximize, where the value of the function for a particular instance of the problem is encoded in a phase.  One would like to manipulate these amplitudes according to their phases, causing the magnitudes of some of the amplitudes to increase and the magnitudes of others to decrease.  In addition, one might wish to learn something about the possible landscape of possible phases, that is how many computational basis states correspond to a given phase.  One natural way to do this is to make use of interference.  Here we wish to investigate how that can be done.

An example of such states are those produced by the QAOA (Quantum Approximate Optimization Algorithm) algorithm \cite{farhi}.  It is a recently proposed quantum procedure for tackling optimization problems.  It has been used to study the MaxCut problem in graph theory \cite{farhi,rieffel1,rieffel2,dam,lukin}, other graph problems \cite{rieffel3}, the Sherrington-Kirkpatrick model \cite{farhi2}, and Boolean satisfiability \cite{montanaro}.  A review of these and other applications can be found here \cite{blekos}.  In this approach, the information about solutions to the problem is encoded into the phases of amplitudes of the computational basis states.  One is then faced with the problem of manipulating and obtaining the information encoded in those phases.  

In an optimization problem one wants to find the computational basis states corresponding to the highest phase.  However, there are other problems one can consider.  Suppose one wishes to produce a state containing only the computational basis states corresponding to a subset of the phases.  Starting with a state containing both amplitudes with phases one wants and phases one doesn't want, constructive interference can be used to enhance the former and destructive interference can be used to decrease the latter.

Let us look in more detail at the application of QAOA to the MaxCut problem for a graph.  In this problem, one is given a graph, $G$, and the object is to divide the vertices into two sets in order to maximize the number of edges between the two sets.  One represents each vertex as a qubit and the state $|0\rangle$ corresponds to being in one set and the state $|1\rangle$ corresponds to being in the other.  We define a Hamiltonian
\begin{equation}
H = \frac{1}{2}\sum_{j,k\in E} (1-\sigma_{j}^{z} \sigma_{k}^{z} ) ,
\end{equation}
where $E$ is the set of edges of the graph and $\sigma_{j}^{z}$ is the $\sigma^{z}$ operator for the $j^{\rm th}$ qubit.  If the qubits corresponding to the vertices $j$ and $k$ are in the same state, the term corresponding to the edge between them is zero, while if they are in different states, the term is one.  Consequently, applying $H$ to a state in the computational basis will simply multiply the state by the number of edges connecting the two sets specified by the state.  We now apply the unitary operator $\exp (-i\alpha H)$ to the state 
\begin{equation}
|\psi_{in}\rangle = \frac{1}{2^{N_{G}/2}} \sum_{x=0}^{2^{N_{G}-1}} |x\rangle ,
\end{equation}
where $N_{G}$ is the number of vertices in the graph and $|x\rangle$ is an $N_{G}$ qubit computational basis state corresponding to the $N_{G}$-digit binary number $x$.  The resulting state, $|\psi_{QAOA}\rangle = \exp (-i\alpha H)|\psi_{in}\rangle$, encodes all possible solutions in the phases of the amplitudes of the computational basis states.  We can choose the parameter $\alpha$ so that these phases lie in the range $[0,\pi ]$.  This is done by choosing $\alpha = \pi /|E|$, where $|E|$ is the number of edges in the graph.  We are interested in finding computational basis states whose amplitudes have phases close to $\pi$, because these correspond to partitions of the vertices with a large number of edges connecting the two sets.  The standard QAOA follows the application of $\exp (-i\alpha H)$ by a mixing unitary, and then iterates this procedure, adjusting parameters in the mixing unitary, to try to find basis sates with large phases.

In the case of QAOA if one has a possible solution, $x$, one can check it classically by using $x$ to find two sets in the graph $G$ and counting the edges that connect them.  This may not be true for other problems.  In particular, if one only has access to the unitary operator $U_{a}|x\rangle = e^{\phi_{x}}|x\rangle$, the checking procedure becomes more difficult requiring the use of a phase estimation protocol.

We will begin by describing the procedure to use measurements and interference to amplify certain phases and discuss how it can be applied to two graph problems, MaxCut and minimum vertex cover.  We will then discuss two examples, and then show how the measurement results can be used to find information about the distribution of phases.  Finally, we will apply the procedure to the QAOA MaxCut state for a number of different graphs.  We find that in the examples we study, there is not a quantum advantage, however, the measure and interference procedure can be used to reduce the number of classical checks that are necessary.

\section{Phase amplification}
We begin with an $N+1$ (for an $N_{G}$ qubit space, $N=2^{N_{G}}-1$) dimensional space, $\mathcal{H}_{a}$, spanned by the orthonormal basis $\{ |x\rangle_{a} \, | \, x =0,1,\ldots N \}$, the state 
\begin{equation}
|\psi_{0}\rangle = \frac{1}{\sqrt{N}} \sum _{x=1}^{N} |x\rangle_{a} ,
\end{equation}
and a unitary operator $U_{a}$, where $U_{a}|x\rangle_{a} = \exp (i\phi_{x}) |x\rangle_{a}$ and $\phi_{0}=0$.  The angles $\phi_{x}$ are between $0$ and $\pi$.  For the MaxCut problem, the states $|x\rangle_{a}$ would correspond to the binary sequences specifying the partition of the graph, and phases are those assigned to each partition by the first step of the QAOA procedure.  We want to increase the amplitude of the states with phases near $\pi$, and decrease the amplitudes of those with a phase near zero.

We now move to the space $\mathcal{H}_{a} \otimes \mathcal{H}_{b}$, where $\mathcal{H}_{b}$ is a copy of $\mathcal{H}_{a}$, and define the operator $U_{ab}$ by
\begin{eqnarray}
U_{ab}(|x\rangle_{a} |0\rangle_{b}) & = & \frac{1}{\sqrt{2}} (|x\rangle_{a}|0\rangle_{b} + |0\rangle_{a}|x\rangle_{b} ) \nonumber \\
U_{ab}(|0\rangle_{a}|x\rangle_{b}) & = & \frac{1}{\sqrt{2}} (|0\rangle_{a}|x\rangle_{b} - |x\rangle_{a}|0\rangle_{b}) .
\end{eqnarray}
We now start with the state $|\psi_{0}\rangle_{a} |0\rangle_{b}$ and apply the operator $U_{ab}(U_{a}\otimes I_{b})U_{ab}$.  The result is
\begin{eqnarray}
|\Psi\rangle_{ab} & = & U_{ab}(U_{a}\otimes I_{b})U_{ab} |\psi_{0}\rangle_{a} |0\rangle_{b}
\nonumber \\
& = & \frac{1}{2\sqrt{N}} \sum_{x=1}^{N} [ (e^{i\phi_{x}} -1) |x\rangle_{a}|0\rangle_{b} \nonumber \\
& & + (e^{i\phi_{x}} +1) |0\rangle_{a} |x\rangle_{b}] .
\end{eqnarray}
We now measure the second system to determine whether it is in the state $|0\rangle_{b}$, i.e.\ we measure the operator $I_{a} \otimes P_{0b} = I_{a}\otimes |0\rangle_{b}\langle 0|$, and keep the result if we get $1$.  The state after a successful measurement is $|\psi_{1}\rangle_{a}|0\rangle_{b}$, where
\begin{equation}
\label{1meas0}
|\psi_{1}\rangle_{a} = \frac{1}{2\sqrt{Np_{1}}} \sum_{x=1}^{N} (e^{i\phi_{x}} -1) |x\rangle_{a} ,
\end{equation}
where
\begin{equation}
\label{p1}
p_{1} = \frac{1}{4N} \sum_{x=1}^{N} | e^{i\phi_{x}} - 1|^{2} ,
\end{equation} 
is the probability for the measurement to succeed.  Note that if we instead get $0$ for the measurement result, which happens with a probability of
\begin{equation}
1-p_{1}= \frac{1}{4N} \sum_{x=1}^{N}  | e^{i\phi_{x}} + 1|^{2} ,
\end{equation}
the state is
\begin{equation}
 \frac{1}{2\sqrt{N(1-p_{1}})} \sum_{x=1}^{N} (e^{i\phi_{x}} +1) |0\rangle_{a} |x\rangle_{b} .
\end{equation}
In that case, we can discard the $a$ system and keep the $b$ system, which contains useful information about the phases, $\phi_{n}$.

Looking at the expression for $|\psi_{1}\rangle_{a}$, we see that the amplitudes for the state with $\phi_{x}$ near $\pi$ have been increased and those for $\phi_{x}$ near $0$ have been decreased.  The process can now be repeated. The success probability for the second measurement is
\begin{equation}
p_{2} = \frac{1}{16Np_{1}} \sum_{x=1}^{N} |e^{i\phi_{x}} -1|^{4}  ,
\end{equation}
and the resulting state will be
\begin{equation}
|\psi_{2}\rangle_{a} = \frac{1}{2\sqrt{Np_{1}p_{2}}} \sum_{x=1}^{N} (e^{i\phi_{x}} -1)^{2} |n\rangle_{a} .
\end{equation}
Let us note three things.  First, the amplification of the phases near $\pi$ and the suppression of the phases near $0$ is even greater for $|\psi_{2}\rangle_{a}$ than for $|\psi_{1}\rangle_{a}$.  Second, since the probability of the high phase states in $|\psi_{1}\rangle_{a}$ is greater than that of the low phase states, $p_{2}$ will generally be greater than $p_{1}$, and the success probability will increase with each iteration.  Third, if we are interested in amplifying not the phases near $\pi$ but those near $\pi - \xi$ for some $\xi$, then we can apply an operator $U_{a}^{\prime}$ right after we apply $U_{a}$, where $U_{a}^{\prime}|x\rangle_{a} = e^{i\xi}|x\rangle_{a}$ for $x\neq 0$ and $U_{a}^{\prime}|0\rangle_{a} = |0\rangle_{a}$.

These expressions can be generalized.  If we have made $m-1$ successful measurements, the success probability for the $m^{\rm th}$ measurement is 
\begin{equation}
\label{prob-gen}
p_{m} = \frac{1}{4^{m}p_{m-1}p_{m-2}\ldots p_{1}N} \sum_{x=1}^{N} |e^{i\phi_{x}} - 1|^{2m} ,
\end{equation}
and the quantum state after the $m^{\rm th}$ successful measurement is
\begin{equation}
\label{state-gen}
|\psi_{m}\rangle = \frac{1}{2^{m}\sqrt{p_{m}p_{m-1}\ldots p_{1}N}} \sum_{x=1}^{N} (e^{i\phi_{x}} - 1)^{m} |x\rangle_{a} .
\end{equation}

These formulas can tell us something useful about applying these techniques to optimization problems.  Suppose we want to find an $x$ with $\phi_{x}>\theta$, for some specified $\theta$ near $\pi$. The idea is to amplify the probabilities of the phases near $\pi$ and then measure the resulting state in the computational basis and find an $x$ with $\phi_{x}\geq \theta$. The probability of making $m$ successful measurements and then measuring in the computational basis and finding an $x$ such that $\phi_{x}>\theta$ is
\begin{equation}
p_{a}(\phi_{x} \geq \theta ) = \frac{1}{N} \sum_{( x\, |\, \phi_{x}\geq \theta )} \left( \frac{1 - \cos\phi_{x}}{2}\right)^{m} .
\end{equation}
If we sampled the original distribution, the one in which each phase, $\phi_{x}$, has the same probability, the probability of finding an $x$ with $\phi_{x}\geq \theta$, $p_{s}(\phi_{x} \geq \theta )$ is
\begin{equation}
p_{s}(\phi_{x} \geq \theta ) = \frac{1}{N} \sum_{( x\, |\, \phi_{x}\geq \theta )} 1,
\end{equation}
and we see that $p_{s}(\phi_{x} \geq \theta ) > p_{a}(\phi_{x} \geq \theta )$, though the difference will not be too great if $\theta$ is near $\pi$..  Therefore, it does not look as though the quantum procedure gives any advantage in finding a good solution to an optimization problem.  However, a closer look reveals that there can be an advantage in some cases.  For example, in the QAOA MaxCut problem what one obtains from sampling the original distribution and the distribution modified by interference and measurement is an $N_{G}$-digit binary number, $x$.  One then has to go back to the graph and check each edge to see whether it connects the two sets corresponding to $x$.  These operations are straightforward and the number of them for a given graph and $x$ is polynomial in the number of vertices of the graph.  However, in both the quantum and the standard sampling procedure the number of times these operations have to be performed is exponential in $N_{G}$.  In the quantum procedure, however, they need to be performed fewer times.  In a subsequent section we will look at an example for a graph with 16 vertices in which the checking operations need to be performed $0.01$ as many times for the quantum procedure as for the standard sampling one.

\section{Minimum vertex cover}
Another graph problem to which this method can be applied is finding the minimum vertex cover.  A vertex cover of a graph is a set of vertices such that every edge in the graph has one or both end vertices in the set.  The minimum vertex cover is the smallest set with this property.  Minimum vertex cover, like MaxCut is an NP complete problem.  There has been work on quantum algorithms for minimum vertex cover.  QAOA algorithms incorporating quantum walks were studied in \cite{Marsh,Wang}, quantum annealing was used in \cite{Hahn}, and variants of QAOA were applied in \cite{Angara,Petruccione}.

In order to apply our interference technique, we need to modify the Hamiltonian.  First, in the $n$-qubit state $|x\rangle$, qubits in the state $|0\rangle$ will correspond to the vertex not being in the cover and qubits in the state $|1\rangle$ will correspond to vertices that are in the cover.  The Hamiltonian will count the number of edges one or both of whose end vertices are in the cover.  It is given by
\begin{eqnarray}
H_{vc} & =  & \sum_{j,k\in E} \left[ \frac{3}{4} - \frac{1}{4}\left(\sigma^{z}_{j}\otimes I_{k} + I_{j}\otimes \sigma^{z}_{k} \right. \right. \nonumber \\
&&\left. \left. + \sigma^{z}_{j}\otimes \sigma^{z}_{k} \right) \right] .
\end{eqnarray} 
Note that if the edge between vertices $j$ and $k$ is such that $|x_{j}\rangle = |0\rangle$ and $|x_{k}\rangle = |0\rangle$, that is neither vertex is in the vertex cover, the expression in brackets is then zero.  For other combinations, $|0\rangle |1\rangle$, $|1\rangle |0\rangle$, and $|1\rangle |1\rangle$, it is one.  We can then apply our amplification procedure as before but with $U_{a}=\exp(-i\alpha H_{vc})$, where $\alpha$ is chosen so that the phases are between $0$ and $\pi$, and $-\alpha$ times the number of edges in the graph is equal to $\pi$.  States corresponding to vertex sets covering a large number of edges will have their amplitudes increased, and then we can sample (measure in the computational basis) from the successfully amplified states.  As before, this will save on the number of checking operations that need to be performed, because these will only be performed on the successfully amplified states.  In this case, the checking operation consists of taking the output, which is one of the states $|x\rangle$ and determining how many of the edges have endpoints in the set to which $|x\rangle$ corresponds.

\section{Examples}
In order to see how the phase amplification works, we will look at examples.  Let us suppose that the phases are evenly distributed, that is, each phase $\phi_{x}$ is equal to $k\pi /N$ for some $k\in \{ 1,2,\ldots N\}$, but we do not know the relation between $k$ and $x$.  We can express the state after the m$^{th}$  successful iteration as 
\begin{equation}
\label{constant-m}
|\psi_{m}\rangle_{a} = \frac{d_{m}}{\sqrt{N}} \sum_{x=1}^{N} (e^{i\phi_{x}}-1)^{m} |x\rangle_{a}, 
\end{equation}
where
\begin{eqnarray}
1 & = & \frac{d_{m}^{2}}{N} \sum_{x=1}^{N} |e^{i\phi_{x}}-1 |^{2m} \nonumber \\
 & = &  \frac{2^{m}d_{m}^{2}}{N} \sum_{x=1}^{N} (1-\cos\phi_{x})^{m} .
\end{eqnarray}
We can approximate the sum by an integral by first rearranging the terms of the sum so that the phases appear in increasing order
\begin{eqnarray}
\frac{1}{N}\sum_{x=1}^{N}  (1-\cos\phi_{x})^{m} & = & \frac{1}{N}\sum_{k=1}^{N}  [1-\cos (\pi k/N)]^{m}
\nonumber \\
 & \simeq & \frac{1}{\pi} \int_{0}^{\pi} d\phi (1-\cos\phi )^{m} .
\end{eqnarray}
We then have that
\begin{eqnarray}
\int_{0}^{\pi} d\phi (1-\cos\phi )^{m} & = & 2^{m-1} \int_{0}^{2\pi} d\phi \sin^{2m}\phi \nonumber \\
 & = & \frac{\pi}{2^{m}} \left( \begin{array}{c} 2m \\ m \end{array} \right) ,
\end{eqnarray}
and, therefore, 
\begin{equation}
d^{2}_{m} =  \left( \begin{array}{c} 2m \\ m \end{array} \right)^{-1} .
\end{equation}

If we start with $|\psi_{m}\rangle_{a}$ and apply our procedure, we find that the probability of a successful outcome, $p_{m+1}$, is 
\begin{eqnarray}
p_{m+1} & = & \frac{2^{m-1}d_{m}^{2}}{N} \sum_{x=1}^{N} (1-\cos\phi_{x})^{m+1}  \nonumber \\
& \simeq & \frac{1}{4}  \left( \begin{array}{c} 2m+2 \\ m+1 \end{array} \right)  \left( \begin{array}{c} 2m \\ m \end{array} \right)^{-1} \nonumber \\
& = & \frac{2m+1}{2m+2} .
\end{eqnarray}
Note that $p_{m}=(2m-1)/(2m)$ is an increasing function of $m$, so that the probability of a successful outcome increases with each iteration.  Let $P_{M}$ be the probability of a sequence of $M$ successful measurements.  We have that
\begin{equation}
P_{M} = \frac{(2M-1)!!}{(2M)!!} = \frac{(2M)!}{2^{2M} (M!)^{2}} .
\end{equation}
Application of the Stirling approximation yields
\begin{equation}
P_{M}\simeq \frac{1}{\sqrt{\pi M}} .
\end{equation}
While this probability is decreasing, it decreases rather slowly.  In order to obtain a sequence of $M$ successful measurements, one would have to try $O(\sqrt{M})$ times.

In our continuum approximation, the probability that a state $|x\rangle$ selected at random will have $\phi_{x} \geq \theta$ (see Eq.\ (\ref{constant-m})) after $M$ successful measurements is 
\begin{equation}
p(\phi_{x} \geq \theta ) = \frac{2^{M}d_{M}^{2}}{\pi} \int_{\theta}^{\pi} d\phi (1-\cos\phi )^{M} .
\end{equation}
As $M$ increases, $(1-\cos\phi )^{M}$ becomes more peaked at $\phi = \pi$.  For $\phi$ close to $\pi$, we have $1-\cos\phi \simeq 2 - (1/2)(\pi - \phi )^{2}$, and setting $\delta = (1/4) (\pi - \phi )^{2}$ gives us $(1 - \cos\phi )^{M} \simeq 2^{M} (1- \delta )^{M}$.  For $\delta$ small, we find 
\begin{equation}
\ln (1-\delta )^{M} = M\ln (1-\delta ) \simeq -M\delta ,
\end{equation}
so that 
\begin{equation}
(1-\cos\phi )^{M} \simeq 2^{M} e^{-M\delta} .
\end{equation}
Inserting this into the expression for $p(\phi_{x} \geq \theta )$ and again making use of the Stirling approximation, for $\gamma$ close to $\pi$
\begin{equation}
p(\phi_{x} \geq \gamma ) \simeq \sqrt{\frac{M}{\pi}} \int_{\theta}^{\pi} d\phi\, e^{-M(\pi - \phi )^{2}/4} .
\end{equation}
We see that if $\pi -\theta \simeq 1/\sqrt{M}$, then this probability is of order one.  So, if we have had $M$ successful measurements, and then measure the state in the computational basis, with high probability we will find a state $|x\rangle$ whose corresponding angle, $\phi_{x}$ is within $1/\sqrt{M}$ of $\pi$.  If we were to sample the original flat distribution, the probability of finding a $\pi - \theta \leq \phi_{x} \leq \pi$, where $\pi -\theta \simeq 1/\sqrt{M}$, is of order $1/\sqrt{M}$.  Therefore one can either sample the original distribution approximately $\sqrt{M}$ times or apply the amplification procedure $\sqrt{M}$ times to produce a state that yields a phase in the desired range when measured in the computational basis.  Note that the checking procedure, which yields the number of edges between the sets, needs to be used $\sqrt{M}$ times for the sampling procedure and roughly once in the amplification procedure.

Next we will look at a case where we modify the distribution of the phases in order to look at only part of it.  Suppose the distribution of the phases $\phi_{x}$ has two peaks, and we want to suppress one and amplify the other.  Let the lower peak be at $\alpha_{l} < \pi /2$, and the upper peak be at $\alpha_{u} > \pi /2$.  In addition, let the fraction of phases in the lower peak be $q_{l}$, and the fraction in the upper peak be $q_{u}$, so $q_{l}+ q_{u} = 1$.  Finally, define $a_{l}=1-\cos\alpha_{l}$, and $a_{u}=1-\cos\alpha_{u}$.  If we make $M$ measurements all with the result $1$, the probability of this happening, which will be denoted $p_{M}$, is (from Eq.\ (\ref{prob-gen}))
\begin{equation}
p_{M} = \frac{1}{2^{M}}( q_{l}a_{l}^{M} + q_{u}a_{u}^{M}) ,
\end{equation}
and the resulting state is (from Eq.\ (\ref{state-gen}))
\begin{eqnarray}
|\psi_{M}\rangle & = & \frac{d_{m}}{\sqrt{N} }\left ( \sum_{\{ x|\phi_{x}=\alpha_{l}\} } (1-e^{i\alpha_{l}})^{M} |x\rangle \right.  \nonumber \\
& & \left. + \sum_{\{ x|\phi_{x}=\alpha_{u}\} } (1-e^{i\alpha_{u}})^{M} |x\rangle \right)  \nonumber  \\
& = & |\psi_{l}\rangle + |\psi_{u}\rangle  ,
\end{eqnarray}
where $d_{m}$ is a normalization constant, $|\psi_{l}\rangle$ is the part of the state for the lower peak and $|\psi_{u}\rangle$ is the part of the state for the upper peak.  The ratio of the upper to the lower peak is 
\begin{equation}
\frac{\|\psi_{u}\|^{2}}{\|\psi_{l}\|^{2}} = \frac{q_{u}a_{u}^{M}}{q_{l}a_{l}^{M}} .
\end{equation}
This implies that if we measure the state in the computational basis after $M$ successful measurements, the probability that we will find a state in the upper peak, $p_{u}(M)$, is
\begin{equation}
p_{u}(M) = \frac{q_{u}a_{u}^{M}}{q_{u}a_{u}^{M} + q_{l}a_{l}^{M}} .
\end{equation}
The probability of obtaining a label $x$ that is in the upper peak by making $M$ successful measurements and then measuring the state in the computational basis is
\begin{equation}
p_{u}(M)p_{M} = \frac{q_{u}a_{u}^{M}}{2^{M}} .
\end{equation}

Let us look at a case where$\|\psi_{u}\|^{2}/\|\psi_{l}\|^{2}$ is small and see what is required to bring it up to a value of $r>1$.  This implies that we want $(a_{u}/a_{l})^{M} = r(q_{l}/q_{u})$ or
\begin{equation}
M=\frac{\log (rq_{l}/q_{u})}{\log (a_{u}/a_{l})} ,
\end{equation}
where, for convenience, we shall assume the logarithms are base $2$.  Note that we can express $p_{M}$ as
\begin{equation}
p_{M}=\frac{a^{M}_{l}q_{l}}{2^{M}} (1+r) .
\end{equation}
As an example, take $q_{u}=1/8$, so $q_{l} = 7/8$, $a_{u}=2$, and $a_{l}=1/8$.  After one successful measurement we have $\|\psi_{u}\|^{2} / \|\psi_{l}\|^{2} = 16/7$ and the probability of this occurring is $p_{1} = 23/128$.  So, starting from a state where the upper peak is $7$ times smaller that the lower one, if we run this procedure approximately 5 times (roughly the number of repetitions we would need to obtain a successful measurement result), we will be able to produce a state in which the upper peak is twice as big as the lower one.  With two successful measurements, for which the probability is $0.128$, we find $\|\psi_{u}\|^{2} / \|\psi_{l}\|^{2} = 36.6$.  That means that with approximately $8$ measurements, we can produce a state that is approximately a superposition of computational basis states whose phases are in the upper peak.  The process is acting as a kind of filter, keeping only states in the upper peak and discarding those in the lower peak.

This process could be useful if we want to obtain information about $|\psi_{u}\rangle$.  We take a collection of systems in the original state, perform the above procedure on them, and the result is a smaller collection of systems that are approximately in the state $(1/\|\psi_{u}\|) |\psi_{u}\rangle$.  Now suppose we want to find out how close $|\psi_{u}\rangle$ is to a symmetric state, i.e.\ what is its overlap with the symmetric subspace of $n$ qubits.  At this point, if $P_{s}$ is the projection onto the symmetric subspace of $n$ qubits, we can just measure $P_{s}$ on each of our systems in the state $(1/\|\psi_{u}\|) |\psi_{u}\rangle$, and if there are $L$ of them, we can obtain an estimate of $\langle P_{s}\rangle$ to an accuracy of approximately $1/\sqrt{L}$. The number of state preparations that will be required for this depends on $q_{u}$, $a_{u}$, and $a_{l}$ to determine the number of successful measurements it would take to approximately eliminate the lower peak, and on $L$, to determine the accuracy to which we want to estimate $\langle P_{s}\rangle$.  Note that there is no dependence on the number of qubits.

Let us compare this to an analogous classical problem.  The unitary operator that determines the phases of the states $|x\rangle$ is analogous to a classical oracle that when fed an $n$-digit binary number $x$ outputs either $0$ if $x$ is in one set, which we shall call the lower set, and $1$ when $x$ is in the other set, which we call the upper set.  We want to know to what extent the upper set is permutation invariant.  We need to estimate for each Hamming weight, $|x|$, the fraction $f_{|x|}=m_{|x|u}/m_{|x|}$, where $m_{|x|}$ is the number of $n$-digit binary numbers with Hamming weight $|x|$ and $m_{|x|u}$ is the number of $n$-digit binary number with Hamming weight $|x|$ in the upper set.  This can be done by choosing $L_{|x|}$ values of $x$ with Hamming weight $|x|$ and seeing how many of them are in the upper set.  This will give us an estimate of $f_{|x|}$ of accuracy approximately $1/\sqrt{L_{|x|}}$.  If we now take $f_{|x|}m_{|x|}$, add up the contributions from each Hamming weight, and divide by the number of values of $x$ in the upper set, we get a number that characterizes the permutation invariance of the upper set (it gives one if the set is permutation invariant), and is the same as the expectation of $P_{s}$ in the quantum case.  The number of uses of the oracle does depend on the number of qubits, because the number of different Hamming weights for which $f_{|x|}$ has to be determined is $n$.


\section{Gaining information about the distribution of phases}
We have seen what happens when we have a sequence of successful measurements.  Now we want to see what kind of information we can gain about the distribution of phases by utilizing measurement sequences in which there is a mixture of successful and unsuccessful measurement outcomes.  What we will find is that the states that are not a result of just successful measurements can be used to none the less gain useful information.

Let us assume that the possible phases are $\theta_{k}=k\pi /M$, where $M$ is the number of edges in the graph and $0\leq k \leq M$, and define $g(\theta_{k})$ to be the number of values of $x$ for which $\phi_{x}=\theta_{k}$.  Note that $\sum_{k=0}^{M} g(\theta_{k}) = 2^{N_{G}}$.  We would like to gain information about $g(\theta_{k})$.  First, let us note that our probabilities can be rephrased in terms of $g$.  For example, 
\begin{equation}
p_{1}= \frac{1}{2N} \sum_{x=1}^{N} (1-\cos\phi_{x}) =\frac{1}{2N} \sum_{k=0}^{M} (1-\cos\theta_{k}) g(\theta_{k}) .
\end{equation}
This equation tells us that if we know $p_{1}$, then we have some information about $g(\theta_{k})$.  If the distribution is flat, that is $g(\theta_{k}) = N/(M+1)$, then
\begin{equation}
p_{1} \simeq \frac{1}{2\pi} \int_{0}^{\pi} d\theta (1-\cos\theta ) = \frac{1}{2} .
\end{equation}
Therefore, if $p_{1}>1/2$, then there are more configurations with phases greater than $\pi /2$ than those with phases less that $\pi /2$.  

We can get more detailed information if we make more measurements.  Let's first look at making two measurements.  Define $F(q,r)$ to be
\begin{equation}
F(q,r) = \sum_{x=1}^{N} (1-\cos\phi_{x})^{q}(1+\cos\phi_{x})^{r} .
\end{equation}
Note that Eq.\ (\ref{p1}) can be expressed as $p_{1}=(1/2N) F(1,0)$.  Now we will make a change of notation.  We want to represent a string of measurements as a string of ones and zeroes, and will let a one correspond to getting $1$ for a result when we measure $P_{0b}$ and zero correspond to getting  $0$ for a result. For one measurement we have the probabilities $p(0) = (1/2N)F(0,1)$ and $p(1)=(1/2N)F(1,0)$.  For two measurements, we have four possibilities, $p(kj)$, where $j$ and $k$ are either one or zero, and $k$ denotes the result of the second measurement, and $j$ the result of the first, so the sequence is written in reverse order.  Now $p(kj)=p(k|j)p(j)$, and from Eq.\ (\ref{1meas0}) we have that 
\begin{eqnarray}
p(1|1) & = & \frac{1}{4Np(1)} F(2,0) \nonumber \\
p(0|1) & = &  \frac{1}{4Np(1)} F(1,1) .
\end{eqnarray}
From this, we obtain $p(11) = (1/4N) F(2,0)$ and $p(01)=(1/4N) F(1,1)$.  Similar results are found for $p(10)$ and $p(00)$.  This suggests the following hypothesis.  Let $y$ be a sequence of zeroes and ones corresponding to a sequence of $m$ measurements.  If $y$ has $q$ ones, then
\begin{equation}
\label{seqprob}
p(y)= \frac{1}{2^{m}N} F(q,m-q) .
\end{equation}
Now we will proceed by induction and assume this is true for $m$, and we want to show it is true for $m+1$.  The normalized state after $m$ measurements is 
\begin{equation}
|\psi_{y}\rangle = \frac{d_{y}}{\sqrt{N}} \sum_{x=1}^{N} (1-e^{\i\phi_{x}})^{q} (1+e^{i\phi_{x}})^{m-q} |x\rangle ,
\end{equation}
where $d_{y}$ is a normalization constant, which satisfies
\begin{equation}
1 = \frac{2^{m}d_{y}^{2}}{N} F(q,m-q)  .
\end{equation}
If we now make one more measurement and obtain $1$, the resulting unnormalized state is
\begin{equation}
|\tilde{\psi}_{1|y}\rangle = \frac{d_{y}}{2\sqrt{N}}\sum_{x=1}^{N} (1-e^{\i\phi_{x}})^{q+1} (1+e^{i\phi_{x}})^{m-q} |x\rangle ,
\end{equation}
and the probability of getting $1$ is just the square of the norm of this state
\begin{equation}
p(1|y) = \frac{2^{m+1}d_{y}^{2}}{4N}  F(q+1,m-q)  = \frac{F(q+1,m-q)}{2F(q,m-q)} .
\end{equation}
Therefore, if we denote the new sequence by $1y$, we have that, using our assumption, that
\begin{equation}
p(1y) = p(1|y)p(y) = \frac{1}{2^{m+1}N} F(q+1,m-q) .
\end{equation}
A similar result is obtained if we had obtained $0$ for our measurement, and this proves our result.  Finally, note that our result can be expressed as 
\begin{equation}
\label{distribution}
p(y) = \frac{1}{2^{m}N} \sum_{k=0}^{M} (1-\cos\theta_{k})^{q} (1+\cos\theta_{k})^{m-q}  g(\theta_{k}) .
\end{equation}

Now let us use some approximate methods to see what the probability $p(y)$, where $y$ is a sequence of zeroes and ones of length $m$, tells us about the distribution of phases.  We will start from Eq.\ (\ref{distribution}).  First, define the function $f_{q,m}(z)=(1-z)^{q}(1+z)^{m-q}$ for $-1\leq z \leq 1$.  The maximum of this function is at $z_{max}=(m-2q)/m$, and setting $\delta z = z - z_{max}$, we have for $|\delta z | \ll 1$, 
\begin{eqnarray}
f_{q,m}(z_{max}+\delta z) & \cong & 2^{m}\left(\frac{q}{m}\right)^{q} \left(\frac{m-q}{m}\right)^{m-q} \nonumber \\
&& \left[ 1 - \left(\frac{m^{3}}{8q(m-q)}\right) \delta z^{2} \right] .
\end{eqnarray}
The width of the peak around the maximum is $[8q(m-q)/m^{3}]^{1/2}$, which reaches a maximum of $\sqrt{2/m}$ when $q=m/2$.  Looking at Eq.\ (\ref{distribution}) we see that $p(y)$ depends on the values of $\theta_{k}$ such that $z_{max} -\sqrt{2/m} \leq \cos\theta_{k} \leq z_{max}+\sqrt{2/m}$.  Define this interval to be $\mathcal{I}_{q,m}$.  If we now define
\begin{equation}
P_{q,m} = \frac{1}{N} \sum_{\theta_{k}\in \mathcal{I}_{q,m}} g(\theta_{k}) ,
\end{equation}
which is just the fraction of values of $\theta_{k}$ that lie in $\mathcal{I}_{q,m}$, we have, approximately
\begin{equation}
p(y) \cong \left(\frac{q}{m}\right)^{q} \left(\frac{m-q}{m}\right)^{m-q} P_{q,m} ,
\end{equation}
so that the probabilities $p(y)$ give us information about $g(\theta_{k})$ in certain regions.  In particular, it gives us a value of $g(\theta_{k})/N$ averaged over an interval of size $1/\sqrt{m}$ centered at $\cos\theta_{k} = z_{max}$.  This implies that the longer the measurement sequences are, the finer the scale at which we probe $g(\theta_{k})$.

\section{Probabilities}
The last relation in the previous section is approximate, but it is also possible to obtain more rigorous relations.  For example, suppose we know $p_{1}$.  What can we say about $P(\phi_{x}>\phi_{r})$, the probability that $\phi_{x}$ is greater than some reference phase $\phi_{r}$?  Now
\begin{eqnarray}
p_{1} & = & \frac{1}{2N}\sum_{x} (1-\cos\phi_{x}) \nonumber \\
& \leq & \frac{1}{2} [ P(\phi_{x}<\phi_{r}) (1-\cos\phi_{r}) + 2P(\phi_{x}\geq \phi_{r})] \nonumber \\
& \leq & \frac{1}{2} [(1-\cos\phi_{r}) + (1+\cos\phi_{r}) P(\phi_{x} \geq \phi_{r}) ] , 
\end{eqnarray}
where we used $P(\phi_{x}<\phi_{r}) + P(\phi_{x}\geq \phi_{r}) = 1$.  From this we have that
\begin{equation}
P(\phi_{x}\geq \phi_{r}) \geq \frac{2p_{1}-(1-\cos\phi_{r})}{1+\cos\phi_{r}} .
\end{equation}
We also have that 
\begin{equation}
p_{1}\geq \frac{1}{2}(1-\cos\phi_{r}) P(\phi_{x}\geq \phi_{r}) ,
\end{equation}
which implies that
\begin{equation}
P(\phi_{x}\geq \phi_{r}) \leq \frac{2p_{1}}{1-\cos\phi_{r}} .
\end{equation}

This result can be generalized.  Suppose $p(11\ldots 1)$ is the probability of measuring $P_{0b}$ and obtaining $1$ $m$ times (see Eq.\ (\ref{seqprob})).  Applying the same reasoning as above, we obtain
\begin{eqnarray}
P(\phi_{x}\geq \phi_{r}) & \geq & \frac{2^{m}p(11\ldots 1) - (1-\cos\phi_{r})^{m}}{2^{m} - (1-\cos\phi_{r})^{m}}  \nonumber \\
P(\phi_{x}\geq \phi_{r}) & \leq & \frac{2^{m}p(11\ldots 1)}{(1 - \cos\phi_{r})^{m}} .
\end{eqnarray}

Now, as an example, in the case that the $\phi_{x}$ are uniformly distributed, we have that $p_{1} =1/2$ and 
\begin{equation}
p(11) = \frac{1}{4\pi} \int_{0}^{\pi} d\phi (1-\cos\phi )^{2} = \frac{3}{8} .
\end{equation}
Substituting these values into the above inequalities we get for one measurement
\begin{equation}
P(\phi_{x}\geq \phi_{r}) \leq \frac{1}{1-\cos\phi_{r}} ,
\end{equation}
and after two measurements
\begin{equation}
P(\phi_{x}\geq \phi_{r}) \leq \frac{3}{2(1-\cos\phi_{r})^{2}} .
\end{equation}
The two-measurement result only produces a better bound than the one-measurement one if $1-\cos\phi_{r} > 3/2$.

We can also obtain useful information for different sequences of measurement results.  In the case that the first measurement is $1$ and the second is $0$, the probability of this happening is
\begin{equation}
p(01) = \frac{1}{4N}\sum_{n=1}^{N} (1-\cos^{2}\phi_{n}) .
\end{equation}
The largest terms in this sum are the ones for which $\phi_{n}$ is close to $\pi /2$.  Suppose we want to get an estimate of the probability that $\phi_{n}$ is between $\pi /2 - \theta$ and $\pi /2 + \theta$, for some $\theta$.  Dividing the sum in the above equation into terms with $|\phi_{n}-\pi /2| \leq \theta$ and terms with $|\phi_{n}-\pi /2| > \theta$, we have that
\begin{eqnarray}
p(01) & \leq & \frac{1}{4} [ (1-P(\pi /2 - \theta < \phi_{n} < \pi /2 + \theta )) \cos^{2}\theta \nonumber \\
&&+ P(\pi /2 - \theta < \phi_{n} < \pi /2 + \theta ) ] ,
\end{eqnarray}
which gives us that
\begin{equation}
P(\pi /2 - \theta < \phi_{n} < \pi /2 + \theta ) \geq \frac{4p(01) - \cos^{2}\theta}{\sin^{2}\theta} .
\end{equation} 
In the case that the $\phi_{n}$ are uniformly distributed, $p(01)=1/8$, and the inequality only gives useful information if $\cos^{2}\theta < 1/2$.

In the case of QAOA these estimates will not prove to be very useful. It is possible to obtain an upper bound on the maximum number of edges between the two sets in MaxCut classically by using a semidefinite relaxation \cite{Nemirovski}.  However, in the case of gaining information about the spectrum of a unitary operator with unknown eigenvalues, they could prove

\section{Examples of graphs}
In order to provide more detailed examples, we will now turn to the MaxCut states for several different graphs and see what happens when we apply our phase amplification procedure.  We will apply it to states resulting from several simple graphs.  In particular, we will look at lines, grids, and a graph consisting of an outer ring of vertices each of which is connected to its neighbors and to a central vertex (star-ring graph). In more detail, the star ring with $q$ vertices has a single central vertex and a set of outer vertices arranged in a ring around it.  Each outer vertex is connected to the central vertex, and adjacent vertices on the ring are connected to each other for a total of $2(q-1)$ edges.  We will focus on the amplification of amplitudes corresponding to large phases.

\begin{figure}[!t]
    \centering
    \subcaptionbox{ Line graph with q vertices}{\includegraphics[width=0.6\columnwidth, height=4.5cm]{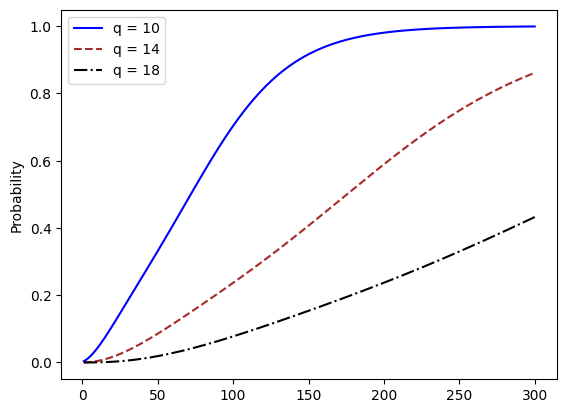}} \\
    \vspace{5mm}
    \subcaptionbox{ Line and grid graphs}{\includegraphics[width=0.6\columnwidth, height=4.5cm]{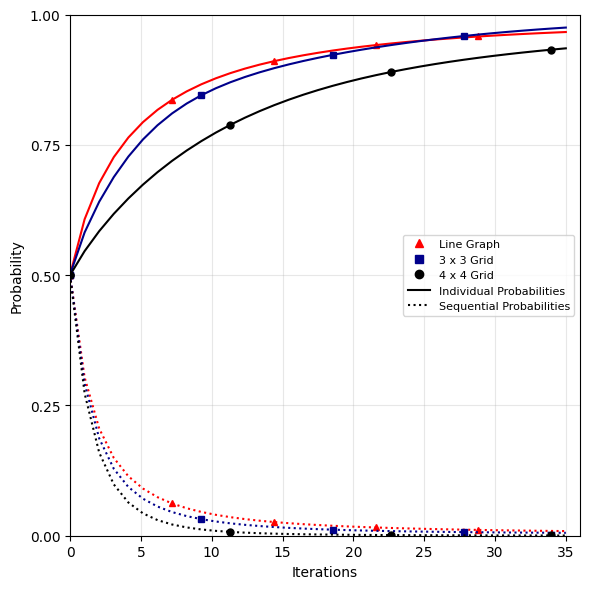}} \\
    \vspace{5mm}
    \subcaptionbox{Grid and star-ring comparison}{\includegraphics[width=0.6\columnwidth, height=4.5cm]{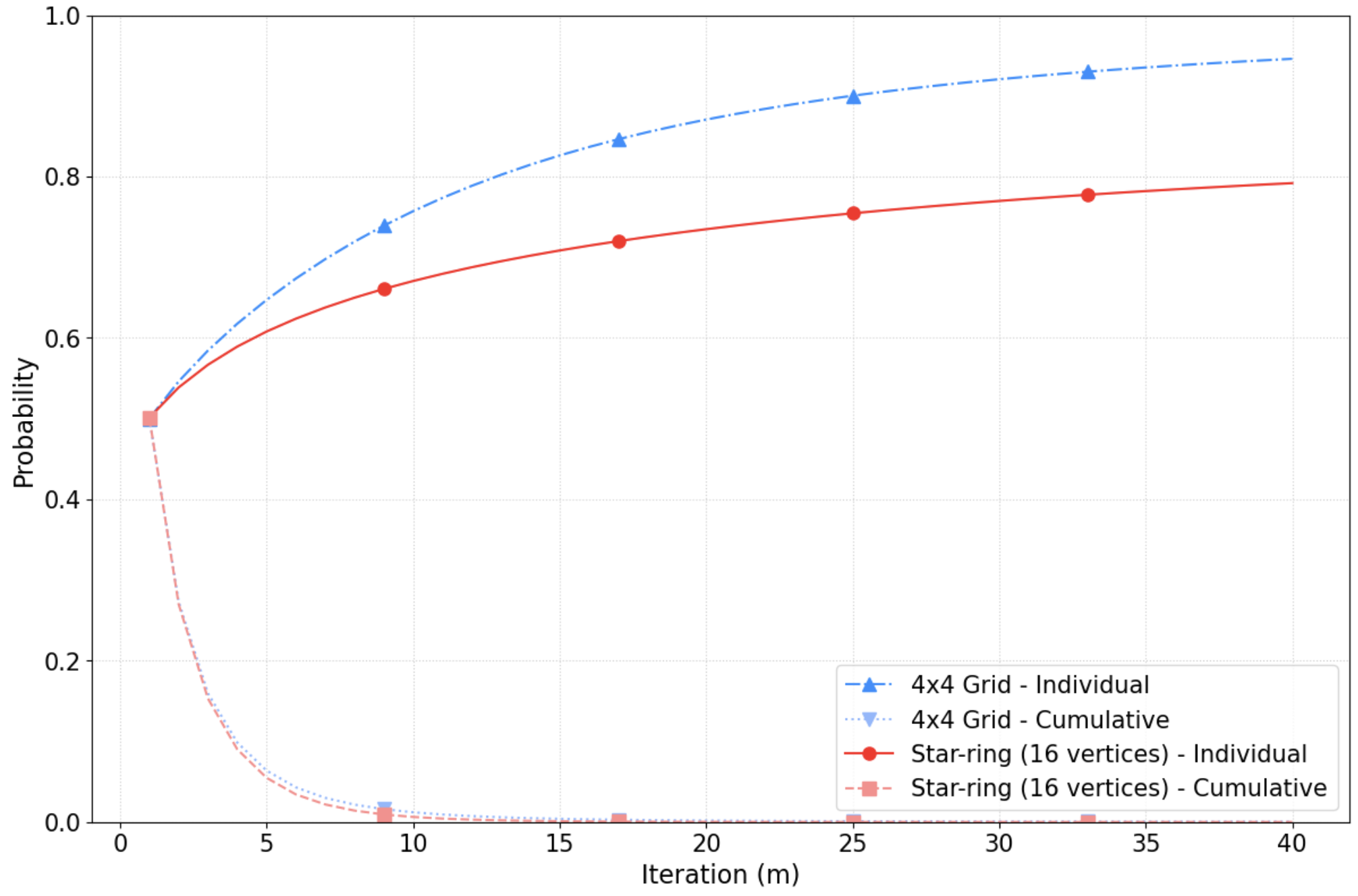}} \\
     \vspace{5mm}
    \caption{Analysis of MaxCut solution probabilities: (a) Shows how total probability of optimal solutions varies for different line graph sizes. (b) Success probabilities for both individual and sequential measurements across graph types. (c) Comparison between grid and star-ring behavior.}
    \label{fig:fig1}
\end{figure}

We will start with lines.  In Fig. 1(a) the sum of the probabilities for the two optimal solutions is plotted versus the number of iterations (successful measurements) for a line graph.  For a line with $q$ vertices and the vertices labeled $\{ 0,1, \ldots q-1\}$, the optimal sets are just $\{ 0,2,4, \ldots q-2\}$ and its complement (we assume $q$ is even). There are two solutions, because each of the two sets can be labeled with zeroes or ones, and there are two ways to do this. The probability being plotted is that of measuring the state in the computational basis and obtaining either of the two optimal solutions.  The state itself is determined by the number of successful iterations of procedure outlined in Sec. II.  As we can see, these probabilities go to one, and the rate at which they do depends on the number of vertices. Fig. 1(b) shows both the sequence and individual probabilities of finding an optimal (or near-optimal) partition versus the number of iterations, for a line with 10 vertices and 3×3 and 4×4 grids. The individual probability is the probability of obtaining 1 when we measure $P_{0b}$.  The sequence probability is the probability of obtaining a given number of successful iterations of the amplification procedure. Initially, there is a steep climb in the individual probabilities, and the rate of increase depends on both the number of vertices and the type of graph. In particular, larger grids require more iterations to achieve probabilities comparable to smaller grids, reflecting their greater complexity and expanded solution space.  We can further see the effect of the shape of the graph on the probabilities in Fig. 1(c). There the individual and sequence probabilities for a $4\times 4$ grid and a star-ring graph with 16 vertices are compared.  We see that the individual probabilities for the grid approach 1 faster than do those for the star-ring.  Both of these graphs have the same number of vertices, and the difference in behavior of the probabilities is only due to the different shapes of the graphs.

We can also gain more insight by looking at some numbers, and we will consider the case of the $4\times 4$ grid. The probability of finding an optimal solution by measuring the initial state of the system in the computational basis is $3.1\times 10^{-5}$.  After $10$ successful iterations the probability of finding an optimal solution by measuring the state is $2.5\times 10^{-3}$.  The probability of obtaining $10$ successful iterations is $0.012$.  In both cases we would have to prepare the initial state $10^{5}$ times.  If we are just measuring the initial state, this task has to be performed $10^{5}$ times in order to have a good chance of finding the optimal solution.  In the second case, we would have to prepare the iterated state approximately $10^{3}$ times to find an optimal solution, and each of these iterated states would require $10^{2}$ preparations of the initial state. Now let us consider what we have to do after we measure the state in the computational basis.  This will give us a partition of the vertices of our graph, and then we have to see how many edges there are between the two parts of the partition. If we measure the initial state, this task has to be performed $10^{5}$ times while if we use the procedure with the iterated state, it only has to be performed $10^{3}$ times.  This suggests that there is an advantage to using the iterated states.

In order to obtain an idea of what the solutions spaces look like, we have plotted the phase distribution function $g(\theta_{k})$ for three different graphs, a line, a grid, and a star-ring, in Fig. 2.  The phase distributions of the line and grid look like a normal distribution, or some slight variation of it. The majority of the states have phases around the midpoint between $0$ and $\pi$ and very few states are near the extreme ends. The ones near $\pi$ correspond to the optimal and near-optimal solutions for the Max-cut problem  (near zero for the Min-cut problem).  The distribution for the star-ring graph is different, and is skewed toward higher phases.  For comparison, we show the probability distribution for the phases for the $4\times 4$ grid after $10$ successful measurements (Fig.\ 3).  This should be compared with Fig.\ 2b.  Note that the higher phases now have increased probabilities compared to the original distribution.  We see then that the space of solutions can look very different for different graphs, and this will have a significant impact on our ability to find good solutions.

\begin{figure}[!t]
    \centering
    \subcaptionbox{ Line graph with 12 vertices}{\includegraphics[width=0.6\columnwidth, height=3 cm]{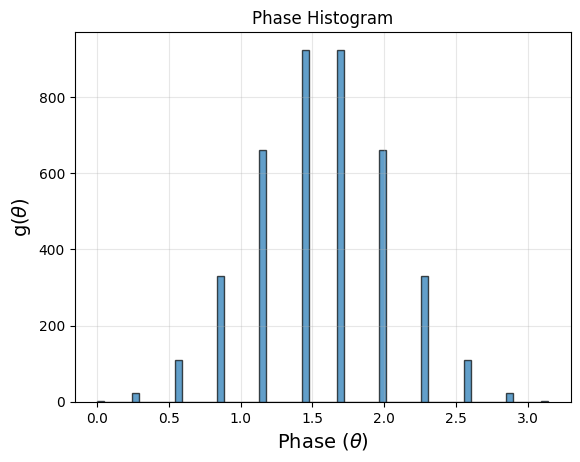}} \\
    \vspace{5mm}
    \subcaptionbox{ Square 4 $\times$ 4 Grid}{\includegraphics[width=0.6\columnwidth, height=3 cm]{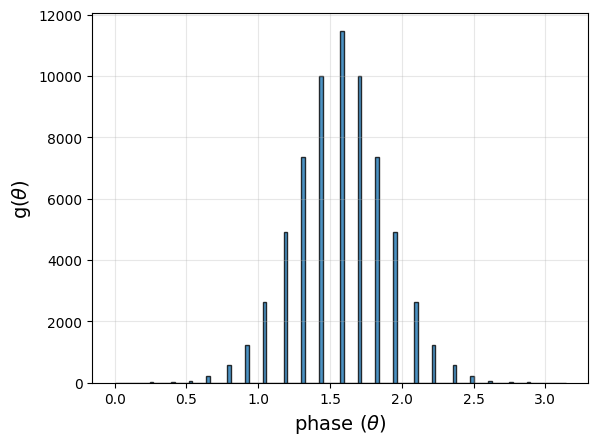}} \\
    \vspace{5mm}
    \subcaptionbox{Star-ring graph with 16 vertices}{\includegraphics[width=0.6\columnwidth, height=3 cm]{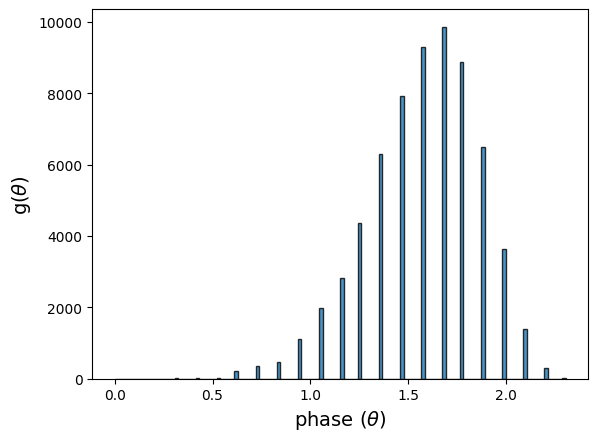}} \\
    \vspace{5mm}
    \caption{Phase distribution of states for a few well-known graphs.}
    \label{fig:phase_dist_all}
\end{figure}
 
 \vspace{5mm}
\begin{figure}[!t]
    \centering
    {\includegraphics[width=0.7\columnwidth, height=3.5 cm]{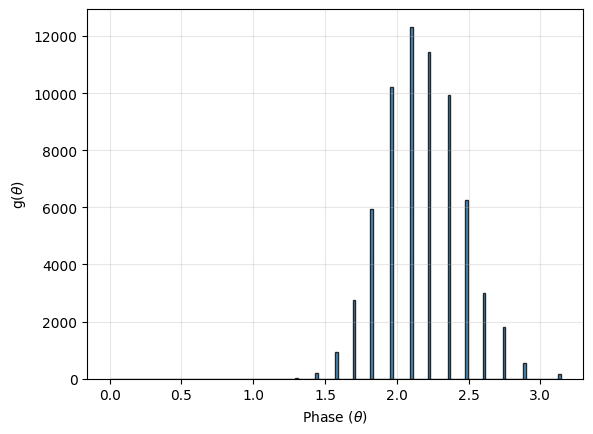}}
    \caption{10th iteration of the amplification procedure on the 4x4 grid graph. $g(\theta)$ here is $2^{16}$ times the probability distribution and is normalized this way for comparison to Fig.\ 2b.  Note the peak has shifted from around 1.5 radians to more than 2 radians, making it likelier to sample a better solution.}
    \label{Fig7.png}
\end{figure}

\section{Conclusion}
We have shown that given a state in which information is encoded in the phases of amplitudes for states in the computational basis interference followed by measurement can be used to amplify certain phases and provide useful information about the distribution of phases in the state.  We looked at two applications of this procedure.  In the first, we implemented the amplification procedure to the state that occurs in the QAOA approach to the MaxCut problem for several graphs, and found that while it does not give a quantum advantage, it can reduce the number of classical checking operations that are required.  In the second, we showed how it is possible to suppress the eigenstates corresponding to part of the spectrum and amplify eigenstates corresponding to other parts.  This allows one to study the properties of a subset of the eigenstates.

\section*{Acknowledgements}
We would like to thank Rudy Raymond for several very useful conversations.  This research was supported by NSF grant FET-2106447.

\end{document}